# Horizontal DPA Attacks against ECC: Impact of Implemented Field Multiplication Formula

Ievgen Kabin
*IHP – Leibniz-Institut für innovative Mikroelektronik*
Frankfurt (Oder), Germany
kabin@ihp-microelectronics.com

Zoya Dyka
*IHP – Leibniz-Institut für innovative Mikroelektronik*
Frankfurt (Oder), Germany
dyka@ihp-microelectronics.com

Dan Klann
*IHP – Leibniz-Institut für innovative Mikroelektronik*
Frankfurt (Oder), Germany
klann@ihp-microelectronics.com

Peter Langendoerfer
*IHP – Leibniz-Institut für innovative Mikroelektronik*
Frankfurt (Oder), Germany
langendoerfer@ihp-microelectronics.com

***Abstract*— Due to the nature of applications such as critical infrastructure and the Internet of Things etc. side channel analysis attacks are becoming a serious threat. Side channel analysis attacks take advantage from the fact that the behavior of crypto implementations can be observed and provides hints that simplify revealing keys. A new type of SCA are the so called horizontal SCAs. Well known randomization based countermeasures are effective means against vertical DPA attacks but they are not effective against horizontal DPA attacks. In this paper we investigate how the formula used to implement the multiplication of *$GF(2^n)$*-elements influences the results of horizontal DPA attacks against a Montgomery *kP*-implementation. We implemented 5 designs with different partial multipliers, i.e. based on different multiplication formulae. We used two different technologies, i.e. a 130 and a 250 nm technology, to simulate power traces for our analysis. We show that the implemented multiplication formula influences the success of horizontal attacks significantly, but we also learned that its impact differs from technology to technology. Our analysis also reveals that the use of different multiplication formulae as the single countermeasure is not sufficient to protect cryptographic designs against horizontal DPA attacks.**

## I. Introduction

The security challenges of today are undergoing a significant change. The Internet of Things, protection of critical infrastructures and the like are requiring a high level of security. For some applications it is not really essential to keep information confidential but to ensure data integrity and authenticity, this holds especially true for monitoring applications in critical infrastructures. The nature of these applications requires that monitoring devices are spread over large areas and are no longer protected by e.g. fences around the premises of their owner. Thus, these devices are physically accessible by potential attackers an issue not taken into account in current security architectures and protocols. In order to avoid malicious attacks the cryptographic implementations need to be protected against side channel analysis (SCA) attacks. SCA attacks exploit the fact that physical effects such as time, power consumption and electromagnetic radiation of cryptographic devices can be measured while performing cryptographic operations. The shape of the measured power traces (PT) or electromagnetic traces (EMT) depends not only on the implemented circuit (technology gates library, area of the design, etc.) but also on the processed input data and the applied cryptographic key. Thus, the measured traces can be analyzed with the goal to reveal the key.

If a single measured trace is not sufficient to extract the key successfully by visual inspection, an attacker can collect many traces. For example, if an elliptic curve cryptography (ECC) design is attacked, the attacker can select specific input data and/or run the device with a selected "key" i.e. a scalar that is processed with the input data in the same manner as the secret key. The collected traces can be analyzed using statistical methods. In order to protect cryptographic implementations against SCA attacks designers try to withhold information from the attacker by blinding the input data, randomization of the key [1] or by randomizing the algorithm steps [2]. Thus, the attacker will no longer reach his/her goal by just altering the input data, since not what the attacker selected is processed but data altered by the implementation. So the attacker can no longer analyze the influence of the data it provided on the shape of the power traces. Meanwhile more sophisticated attacks e.g. those described in [3] and [4] allow to extract secret keys even in case countermeasures are applied.

It is important to note that well-known countermeasures proposed in [1] are effective against collision based [5]-[7] and vertical DPA [8] attacks against ECC designs. But these countermeasures are not effective against horizontal DPA [8] attacks, i.e. against attacks analysing only one measured trace using statistical methods, for example using the difference of the mean as described in [9]. Even if each key bit is processed using exactly the same operation sequence, the data dependability can result in a successfully revealed key if horizontal DPA attacks are applied [10]. Blinding of the elliptic curve (EC) point *P* or randomization of the projective coordinates of point *P* once at the beginning of the *kP* calculation do not provide protection against horizontal DPA attacks [11]. The randomization of the private key *k* does not hinder horizontal DPA attacks because the revealed randomized key can successfully be used instead of the real private key. This means that implementations protected against vertical DPA by such randomization are not implicitly protected against horizontal DPA.

Vertical DPA attacks need more measurements than horizontal attacks to be successful i.e. they are more complex and time consuming. Thus, protection against vertical DPA is only useful if the design is well protected against horizontal DPA attacks. In comparison to vertical DPA attacks, horizontal DPA attacks are relative simple but nevertheless they require by far more complex kinds of randomization to ensure that the randomization is an effective countermeasure.

In this paper we investigate the influence of the implemented field multiplication formula on the success of horizontal DPA attacks. If the multiplication method (MM) is selected properly it does not increase the execution time of the *kP* operation but can increase the resistance against horizontal attacks and thus can be used for efficient implementations of ECC designs.

The rest of this paper is structured as follows. In section II we explain the implementation details of the investigated designs. In section III we present in detail how we performed the horizontal DPA attack using the difference of the mean method. In section IV we discuss the results of the performed attacks. The paper finishes with short conclusions.

## II. Investigated ECC Designs

### A. Structure of the implemented kP designs

All implemented ECC designs are hardware accelerators for elliptic curve point multiplication and have the same structure (see Fig.1).

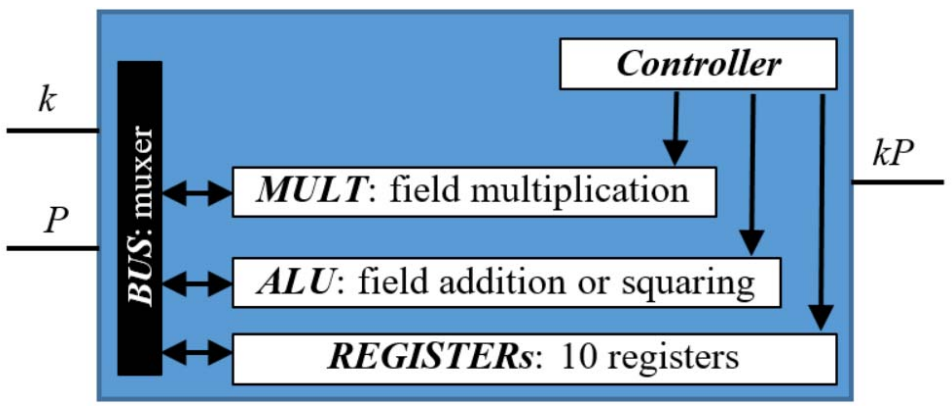

Fig. 1. Structure of the implemented *kP*-accelerators.

The *kP* accelerator obtains a scalar *k* and two affine coordinates *x* and *y* of a point *P* of the EC B-233 [12] as inputs to process. The numbers *x*, *y* and *k* are up to 233 bit long binary numbers that represent elements of $GF(2^{233})$, with the irreducible polynomial $f(t)=t^{233}+t^{74}+1$. The hardware accelerator processes the scalar *k* bitwise according to the Montgomery *kP* algorithm using Lopez-Dahab projective coordinates [13]. In our implementation the main loop of the algorithm consists of multiplications, squarings and additions of the $GF(2^{233})$-elements. The division of $GF(2^{233})$-elements is calculated as a sequence of field squarings and field multiplications using the little theorem of Fermat: $x(t)^{-1} \bmod f(t) = x(t)^{2n-2} \bmod f(t)$

The block MULT calculates the field product. The block ALU performs addition or squaring of its operands depending on the signals of the block Controller. Additionally, the design contains registers for saving intermediate values. The block Controller manages the sequence of the field and write to register operations. It controls the data flow between the blocks and defines which operation has to be performed in the current clock cycle.

The implemented algorithm is a modification of the Montgomery *kP* algorithm that is based on Algorithm 2 in [9]. This implementation allows to perform all operations in parallel to the field multiplications. This increases the efficiency of the design and provides some kind of robustness against simple SCA attacks, for example against SPA and SEMA. In our implementation the processing time of a key bit is equal to the time needed to execute 6 field multiplications.

### B. Field multiplier

The multiplication is the most complex field operation in our designs. The polynomial multiplication (i.e. the first step of the multiplication of elements of $GF(2^n)$) can be realized by applying the classical multiplication method. Its gate complexity (GC) can be given as a number of Boolean AND and XOR operations, i.e. as the number of used AND and XOR gates. To implement the multiplication of *n*-bit long polynomials using the classical multiplication method $n^2$ AND and $(n-1)^2$ XOR gates are necessary. This results in an expensive implementation with respect to area and energy since the length of multiplicands is typically large (about 200 bit). In order to tackle this complexity issue many optimizations, i.e. new multiplication formulae, have been proposed in the past. Many multiplication methods apply segmentation of both multiplicands into the same number of parts. The product is then calculated as a sum of smaller partial products. Historically, the first optimization was the Karatsuba multiplication method published in 1962 [14]. This method uses the segmentation of polynomials into two terms. The next multiplication formula was proposed by Winograd in 1980 [15]. This method uses the segmentation of polynomials into three terms. At the moment there exist a lot of multiplication methods exploiting different multiplication formulae or their combinations. Multi-segment-Karatsuba MM (MSK) [16] and enhanced MSK [17] are examples of such combinations. In [18] and [19] different multiplication methods were combined with the goal to find the optimal combination, i.e. the combination with minimal gate complexity and energy consumption. Each combination of multiplication methods (MM) has its own gate complexity. Additionally the number of XOR gates can be significantly reduced by calculating the product iteratively [20]. Combinations of multiplication methods with reduced XOR-complexity are investigated in [21] for different length of operands and their segmentation.

We implemented the field multiplier using the 4-segment Karatsuba multiplication method according to a fixed calculation plan as described in [20]. The structure of our field multiplier is shown in Fig.2.

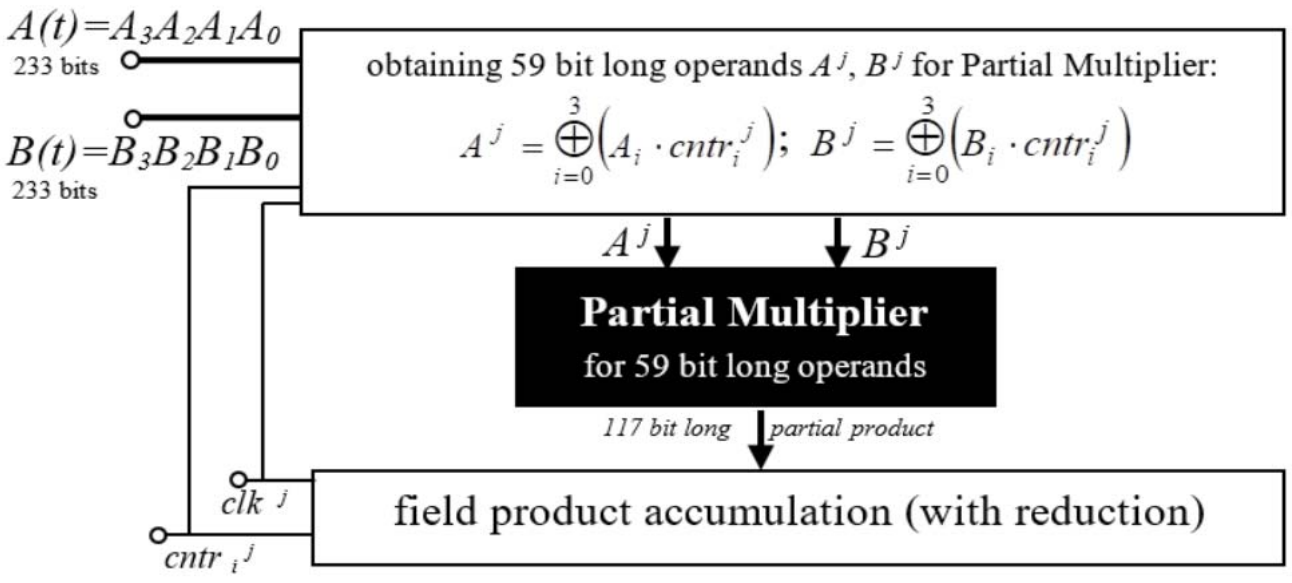

Fig. 2. Structure of the field multiplier.

Two up to 233 bit long operands *A(t)* and *B(t)* are segmented into four parts: $A_3$, $A_2$, $A_1$, $A_0$ and $B_3$, $B_2$, $B_1$, $B_0$ respectively. The parts $A_3$ and $B_3$ are 56 bit long. All other parts are 59 bit long. The field multiplier takes 9 clock cycles ($clk^j$, $j=0,1,2,...,8$) to calculate the product of 233 bit long operands. In each clock cycle only one of 9 partial products of the 59 bit long operands is calculated accordingly to the 4 segment iterative Karatsuba MM.

The signals from the Controller $cntr^i_j$ organize the calculation of operands for the Partial Multiplier clockwise. For example; in the first of the nine clock cycles $clk^0$ of a field multiplication the signals from the Controller are: $cntr^0_0=1$; $cntr^0_1=0$; $cntr^0_2=0$; $cntr^0_3=0$ and the operands for the Partial Multiplier are:

$$A^0 = \bigoplus_{i=0}^{3}\left(A_i \cdot cntr_i^0\right) = A_0 \cdot cntr_0^0 \oplus A_1 \cdot cntr_1^0 \oplus A_2 \cdot cntr_2^0 \oplus A_3 \cdot cntr_3^0 = \\ = A_0 \cdot 1 \oplus A_1 \cdot 0 \oplus A_2 \cdot 0 \oplus A_3 \cdot 0 = A_0 \quad , \tag{1}$$

$$B^0 = \bigoplus_{i=0}^{3}\left(B_i \cdot cntr_i^0\right) = B_0 \cdot cntr_0^0 \oplus B_1 \cdot cntr_1^0 \oplus B_2 \cdot cntr_2^0 \oplus B_3 \cdot cntr_3^0 = \\ = B_0 \cdot 1 \oplus B_1 \cdot 0 \oplus B_2 \cdot 0 \oplus B_3 \cdot 0 = B_0 \quad . \tag{2}$$

All partial products are accumulated in a register of the multiplier. The field product will be accumulated iteratively, step-by-step (or clock-by-clock), using the calculated partial products. The reduction is also performed in each clock cycle. The reduction can be performed only once, after all 9 partial multiplications, but the reduction consumes energy, i.e. it is a kind of "dummy operation" that hides partially the activity of other blocks of the *kP* design. Thus, performing the reduction clockwise can increase the resistance of the *kP* design against SCA attacks, but we did not yet evaluate this effect. Our five designs investigated here differ only in the block Partial Multiplier. In the next subsection we explain the implementation details of our partial multipliers and give their parameters such area and average power consumption.

### C. *Implemented partial multipliers*

*1)* Partial multiplier PM_1.

This partial multiplier was implemented using the classical multiplication formula only, i.e. it implements the following formula:

$$C = A \cdot B = \sum_{i=0}^{2n-2} c_i \cdot t^i, \quad \text{with} \quad c_i = \bigoplus_{i=k+l} a_k \cdot b_l, \quad \forall k,l < n \tag{3}$$

Here $n$=59 is the length of the partial multiplicands.

The gate complexity of this multiplier, i.e. the amount of AND and XOR gates which are necessary to implement its functionality corresponding to formula (3) is $n^2$ of AND gates and $(n-1)^2$ of XOR gates, i.e.: $GC_{n=59}=3481_{\&}+3364_{XOR}$.

*2)* Partial multiplier PM_2.

This partial multiplier of 59 bit long operands was implemented using the 2-segment iterative Karatsuba multiplication formula [21] for 60-bit long multiplicands. The gate complexity of this multiplier is $GC_{2m}=3 \cdot GC_m+(7m-3)_{XOR}$. Here $m$ is the length of segments $m=60/2=30$ and $GC_m$ is the gate complexity of the internal *m-bit* partial multipliers. In PM_2 we implemented all 3 internal 30-bit partial multipliers identically. Each of them was implemented using the 6-segment iterative Winograd multiplication formula [21], with a gate complexity of: $GC_{6m}=18 \cdot GC_m+(72m-19)_{XOR}$, with $m=30/6=5$ bits. Corresponding to the 6-segment iterative Winograd multiplication formula the 30-bit multiplier consists of 18 internal multipliers of 5-bit long operands. Each of these small multipliers was implemented using the classical multiplication formula (3) with $n=5$. Fig.3 shows the structure of PM_2.

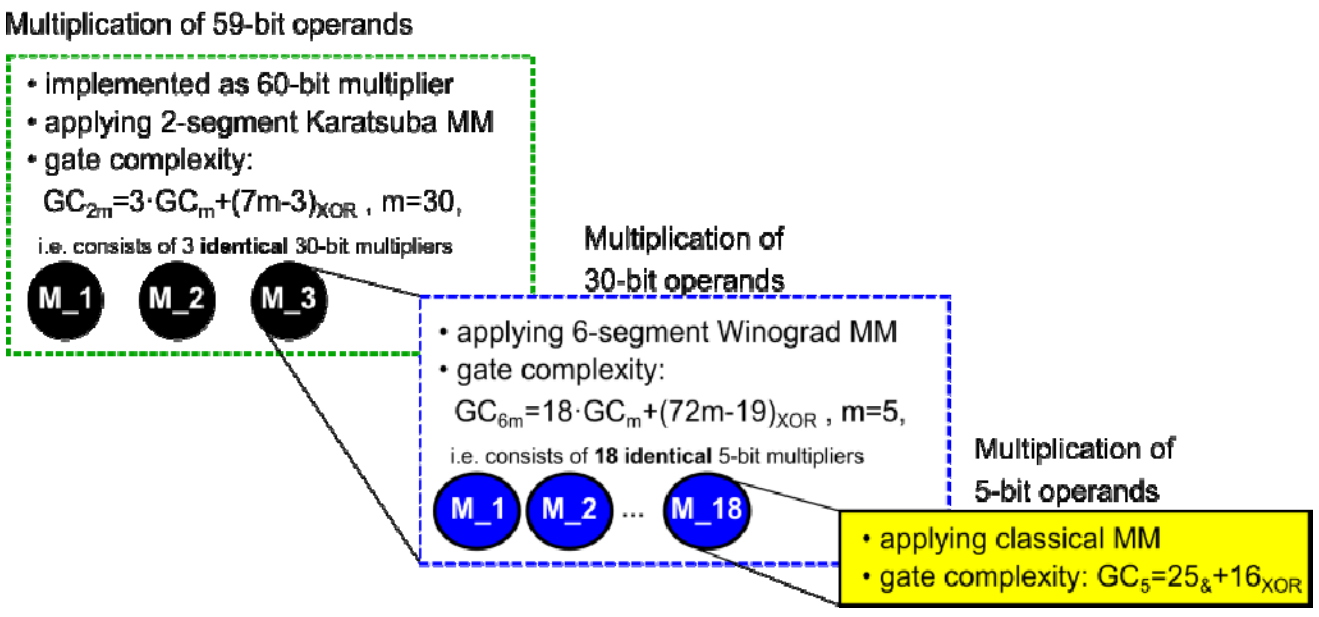

Fig. 3. Partial multiplier *PM_2*.

*3)* Partial multiplier PM_3.

This partial multiplier of 59 bit long operands was implemented using the 4-segment iterative Karatsuba multiplication formula for 60-bit long multiplicands. The gate complexity of this multiplier is $GC_{4m}=9 \cdot GC_m+(34m-11)_{XOR}$. Here $m$ is the length of the segments $m=60/4=15$ and $GC_m$ is the gate complexity of the internal *m-bit* partial multipliers. In PM_3 we implemented all 9 internal 15-bit partial multipliers identically. Each of them was implemented using the 4-segment iterative Karatsuba multiplication formula again. Thus, the 15-bit multiplier consists of 9 internal 4-bit multipliers, each implemented using the classical multiplication formula (3) with $n=4$. Fig.4 shows the structure of PM_3.

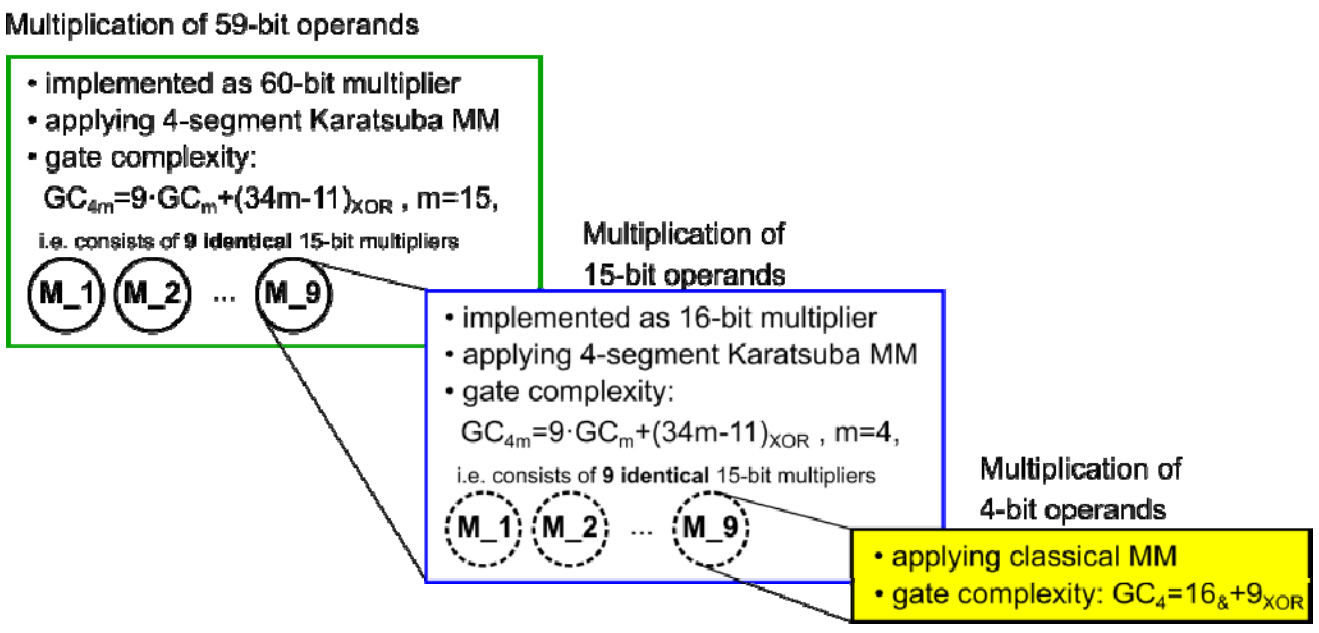

Fig. 4. Partial multiplier *PM_3*.

*4)* Partial multiplier PM_4 and partial multiplier PM_5.

These Partial Multipliers of 59 bit long operands were implemented as a combination of 3 multiplication formulae: the 3-segment iterative Winograd MM, the 4-segment iterative Karatsuba MM and the classical MM. At first the 3-segment iterative Winograd MM was applied for 60-bit long operands. This MM defines the number of the internal multipliers: PM_4 and PM_5 contain 6 internal multipliers *M_1,...,M_6* for 20-bit long operands. The multipliers *M_1*, *M_5* and *M_6* in *PM_4* are identical (see blue marked multipliers in Fig.5-*a)*). The multiplier *M_5* and *M_6* in *PM_5* are identical to the multiplier *M_1* in *PM_4* and are implemented using the classical MM only (see yellow marked multipliers in Fig.5-*a)* and *b)*). The combination of MMs for the implementation of the multipliers $M_1,...,M_4$ in PM_4 and all multipliers of *PM_5* was chosen randomly.

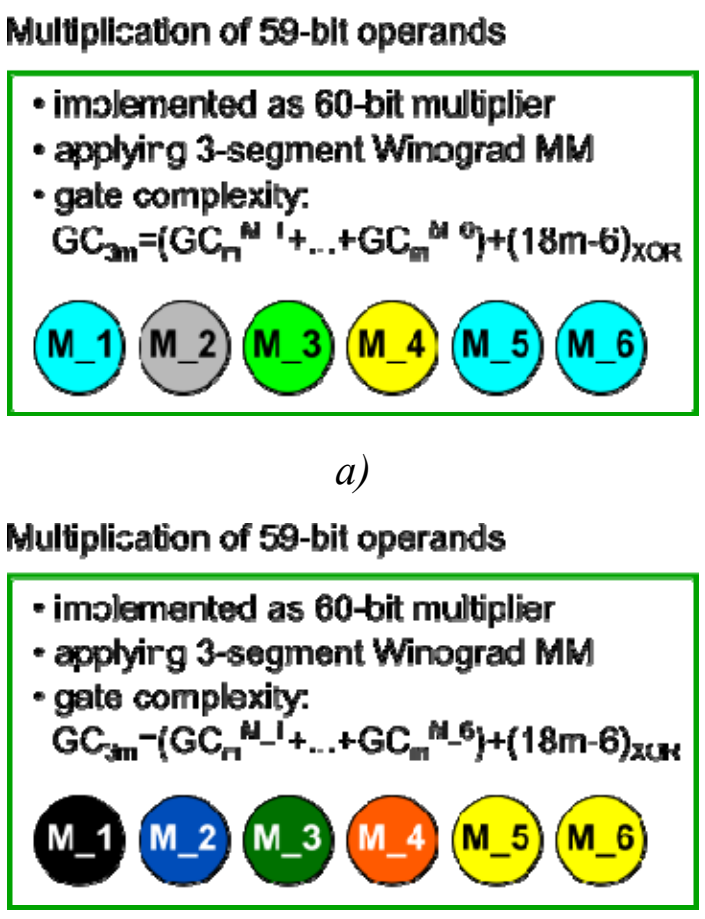

Fig. 5. Partial multipliers implemented using a combination of different multiplication formulae: *a) - PM_4*; *b) – PM_5*.

Details about the iterative 4-segment Karatsuba MM and the 3-segment Winograd MM are given in [22] and [19]. We do not give details here for the simplifying the reading. Important is only the fact, that the complexity of the PMs is different.

We synthesized the *kP* designs using the gate library for our 130 nm and 250 nm technologies for a clock cycle of 30 ns. All investigated designs require the same amount of clock cycles (about 14000) for a single *kP* operation. We simulated the power consumption of the *kP* designs while the *kP* operation was executed. All *kP* traces were simulated using the Synopsis PrimeTime suite [23]. In the rest of this paper the information about the used technology is given as a part of the name of the simulated power trace.

Table 1 gives a short overview of the parameters of the investigated *kP* designs, which use the here described partial multipliers. The goal is to show that the main parameters of the investigated *kP* designs such as their area and power are different.

TABLE I. PARAMETERS OF INVESTIGATED *KP* DESIGNS

| Technology | Investigated *kP* designs | | | Partial multiplier | | | | |
|---|---|---|---|---|---|---|---|---|
| | | | | | area | | power | |
| | *name* | *area, mm²* | *power mW* | *name* | *total mm²* | *relative to kP design, %* | *total mW* | *relative to kP design, %* |
| *130 nm* | *design1* | 0.31 | 7.8 | PM_1 | 0.14 | 45.8 | 5.49 | 70.8 |
| | *design2* | 0.27 | 5.4 | PM_2 | 0.11 | 39.2 | 3.13 | 57.9 |
| | *design3* | 0.28 | 5.6 | PM_3 | 0.11 | 39.8 | 3.33 | 59.5 |
| | *design4* | 0.29 | 6.0 | PM_4 | 0.13 | 43.1 | 3.74 | 62.3 |
| | *design5* | 0.29 | 6.1 | PM_5 | 0.13 | 43.0 | 3.79 | 62.5 |
| *250 nm* | *design1* | 1.50 | 51.1 | PM_1 | 0.61 | 41.0 | 32.6 | 63.7 |
| | *design2* | 1.38 | 40.6 | PM_2 | 0.50 | 36.0 | 22.1 | 54.3 |
| | *design3* | 1.40 | 42.7 | PM_3 | 0.51 | 36.7 | 24.2 | 56.5 |
| | *design4* | 1.46 | 45.1 | PM_4 | 0.58 | 39.5 | 26.6 | 58.8 |
| | *design5* | 1.46 | 46.0 | PM_5 | 0.57 | 39.4 | 27.4 | 59.6 |

## III. PERFORMED HORIZONTAL DPA ATTACK

The simulated traces are noiseless and no data are lost in simulations. Due to this fact we represented each clock cycle using only one power value – i.e. the average power value of the clock cycle. This is reasonable, because it simplified the statistical analysis of the trace without any loss of information relevant for the analysis. Fig. 6 shows a part of each PT and the processed key bit values.

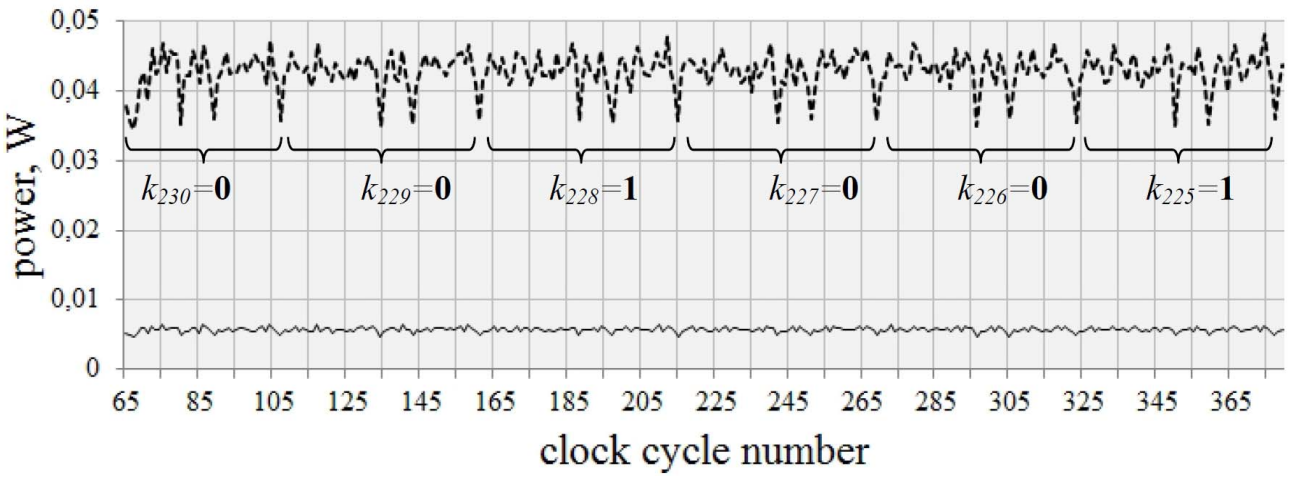

Fig. 6. A part of the power traces simulated for the execution of the *kP* operation. The graph shows power (in W) over clock cycles. The solid line is the PT simulated for the 130 nm technology; the dotted line represents the PT simulated for the 250 nm technology. The processing of the following 6 key bits is shown: $k_{230}=0$, $k_{229}=0$, $k_{228}=1$, $k_{227}=0$, $k_{226}=0$ and $k_{225}=1$. The duration of the processing of the key bit $k_{230}$ is 45 clock cycles. This key bit is not processed in the main loop of the Montgomery *kP* algorithm with the goal to avoid vulnerability to SPA [24]. The duration of each other slot is 54 clock cycles.

We decided to apply the *difference of the mean* for statistical analysis of the traces. To perform a horizontal DPA attack we fragmented the compressed trace into slots. The compressed trace consists of $l\text{-}2=230$ slots with similar profiles, where $l$ is the length of the processed scalar $k$. Each slot corresponds to the processing of a key bit $k_i$ with $0 \le i \le l\text{-}3$ in the main loop of the implemented algorithm and consists of 54 values (one value per clock cycle). Thus, each value of the compressed trace can be represented as $v_i^j$, where $i$ is the number of the slot and $j$ is the number of the clock cycle within the slot, with $0 \le i \le l\text{-}3$, $1 \le j \le 54$. We calculated the *mean* slot, i.e. the arithmetical mean of all values with the same number $j$ and different number $i$: $\overline{v^j} = \frac{1}{l-2}\sum_{i=0}^{l-3} v_i^j$ .

Thus, the 54 calculated values $\overline{v^j}$ define the *mean* slot. We obtained 54 key candidates – one per clock cycle $j$ – using the following assumption: $k_{candidate}{}^j{}_i = 1$ if $\overline{v^j} \ge v_i^j$ else $k_{candidate}{}^j{}_i = 0$. To evaluate the success of the attack we compared the extracted key candidates with the scalar $k$ that was really processed. For each key candidate we calculated its relative correctness as follows:

$$\delta_1 = \frac{\#correct_extracted_bits(k_{candidate}{}^j{}_i)}{l-2} \cdot 100\% \qquad (4)$$

A correctness close to 0 percent means that our assumption is wrong and the opposite assumption will be correct. Taking this fact into account we can calculate the correctness as a value between 50 and 100% as follows:

$$\delta = 50\% + |50\% - \delta_1| \qquad (5)$$

Fig.7 shows the calculated correctness of the key candidates obtained by analysing the traces *design2_130nm* and *design2_250nm* as a line. The solid black line shows the results of the performed horizontal attack using the power trace *design2_130nm*. The dotted black line represents the results obtained using trace *design2_250nm*.

As a reference we defined the worst-case of the attack result from the attacker's point of view as 50 per cent which means that the difference of the mean method cannot even provide a slight hint whether the key bit processed is more likely a '1' or a '0', i.e. this means that the attack was not successful at all. The worst-case from the attacker's point of view is the ideal case from the designer's point of view. We denote it as the "ideal case" in the rest of the paper and represented it in Fig.7 as the green line.

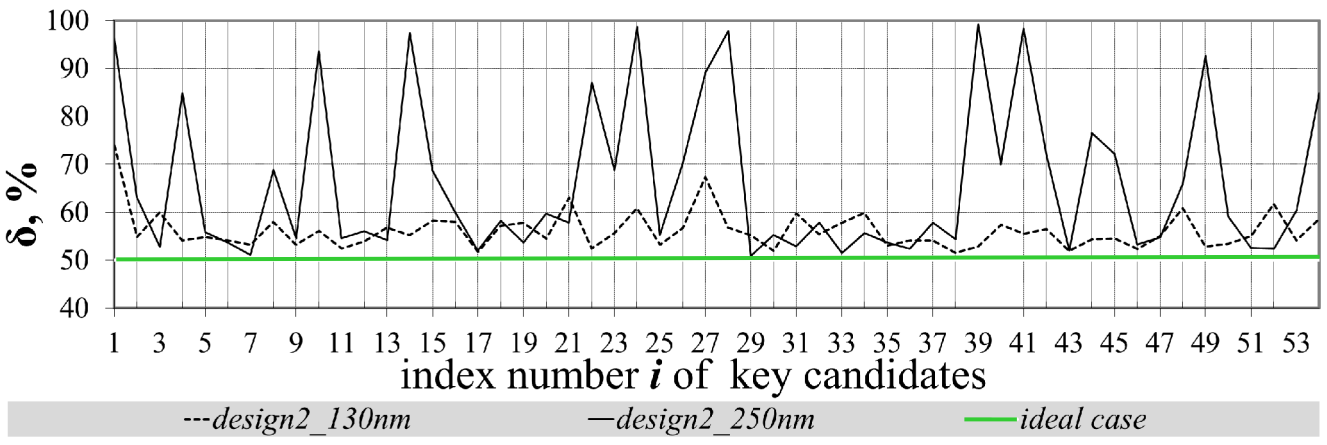

Fig. 7. Horizontal attack using *the difference of the mean method*: the analysis results are obtained using the trace of the *kP* design synthesized for the 130 nm (dotted line) and 250 nm (solid line) technologies.

Fig.7 shows that the *kP* design, that was synthesized for the 130 nm technology, is more resistant against the performed attack than the design synthesized for the 250 nm technology. Only four key candidates – $k_{candidate}^{1}$, $k_{candidate}^{21}$, $k_{candidate}^{27}$ and $k_{candidate}^{52}$ – were extracted with a relative high (more than 60%) correctness of 74%, 63%, 67% and 62% respectively (see doted black line). The correctness of the other 50 key candidates is in the range between 50% and 60%.

Although we implemented the *kP* design strongly balanced and the field multiplier is always active, the key was revealed successfully using statistical methods. The analysis of the power trace *design2_250nm* reveals that 20 of 54 key candidates have a correctness between 70% and 100% (see black dotted line in Fig.7). The information leakage source in our *kP* design is the bus activity. We learned this by analysing the power traces of single blocks of the *kP* design. The relative power consumption of the bus is higher than in the *kP* design synthesized for the 130 nm technology. Thus, the activity of the bus wasn't hidden by the activity of the other blocks of the *kP* design.

## IV. Analysis Results

Fig. 8 and Fig. 9 show the results of the attacks for all investigated traces. For both technologies the *kP* design with the partial multiplier that implemented the classical multiplication formula (see traces design1_130nm and design1_250nm) is the best i.e. the correctness of the extracted key candidates for this design is significantly smaller than for all other designs. Fig. 10 *a)* and *b)* show the correctness of key candidates given in Fig. 8 and Fig. 9 as bars. Key candidates are sorted from the highest to the lowest value of correctness. This representation helps to compare the results of the attack.

The results of the analysis show that the implemented formula for the partial multiplication has a significant impact on the resistance of the *kP* design against horizontal DPA. The *kP* design with the partial multiplier that implemented the classical multiplication formula shows the highest resistance. This is due to the fact that not only the average power consumption of designs with the classical partial multiplier is the highest one, but also the amplitude range of the power trace, in comparison to other designs investigated here.

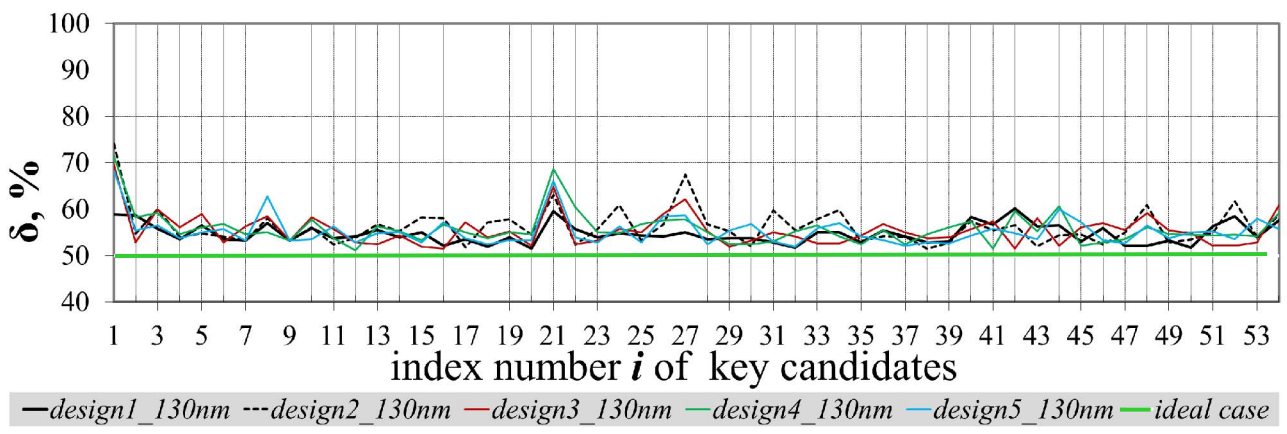

Fig. 8. Horizontal DPA attack using *the difference of the mean* method: the analysis results are obtained using the traces of the *kP* design synthesized for the 130 nm technologies. The *kP* design with the PM that was implemented using the classical multiplication formula is the best: no key candidates are extracted with the correctness higher than 60% (solid black line).

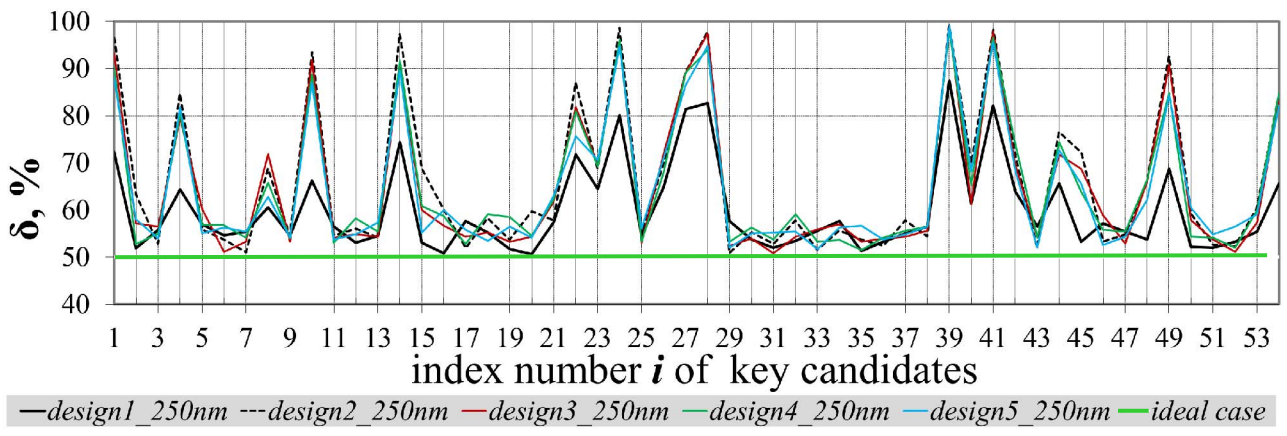

Fig. 9. Horizontal DPA attack using *the difference of the mean* method: the analysis results are obtained using the traces of the *kP* design synthesized for the 250 nm technologies. The correctness of revealing the processed scalar is very high in all investigated cases, but the *kP* design with the Partial Multiplier that was implemented using the classical MM is the best (solid black line).

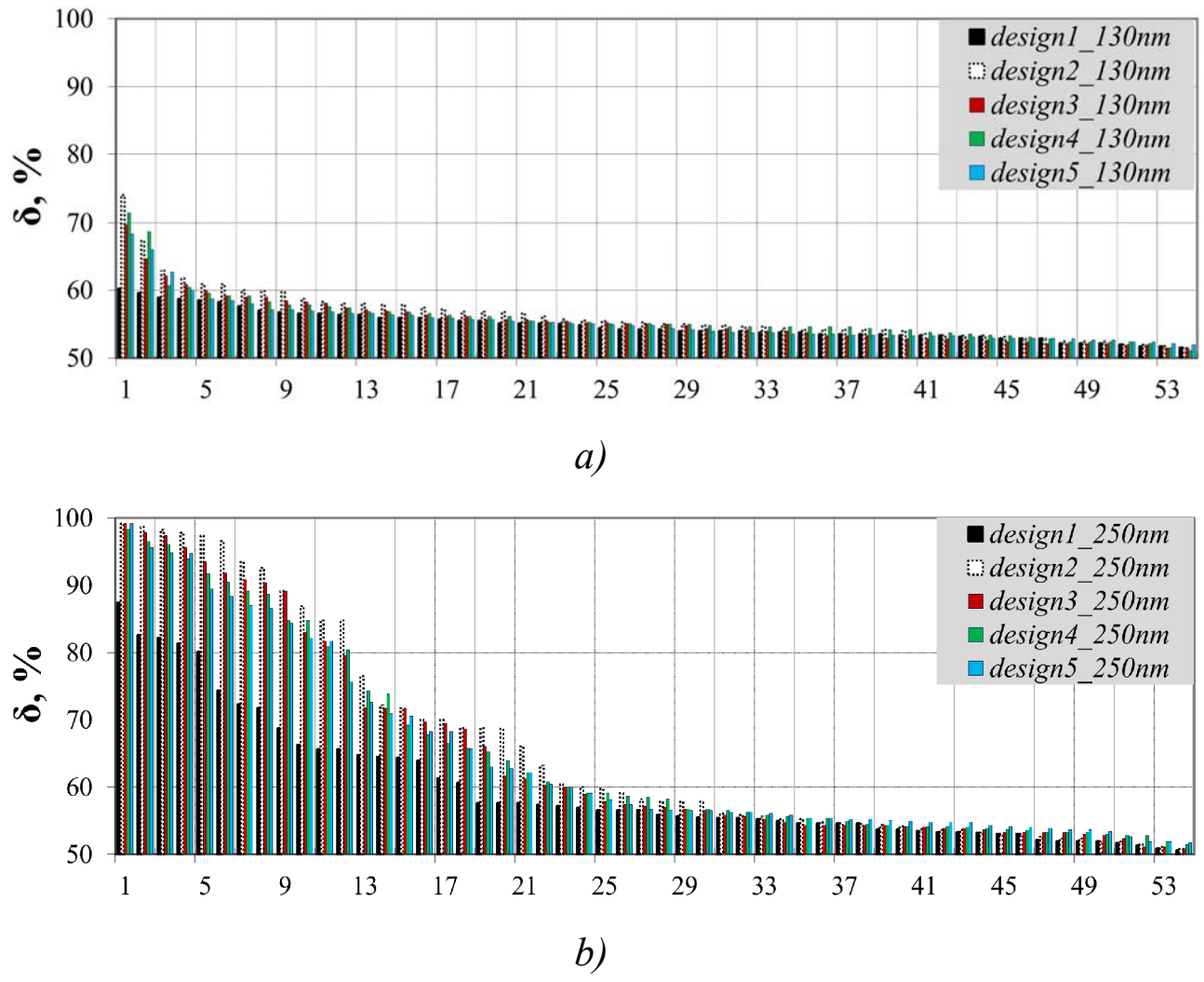

Fig.10. Horizontal DPA attack using *the difference of the mean* method: all key candidates were sorted from the highest to the lowest value of correctness; results for two technologies. Here it can be clearly seen that the design with the Partial Multiplier that was implemented using the classical MM has the lowest correctness of the key extraction in comparison to other designs.

## V. Conclusions

In this paper we investigated the impact of the implemented formula for the partial multiplication on the results of a horizontal DPA attack. We investigated 5 *kP* designs synthesized for two different technologies. The designs differ only in their partial multipliers: each multiplier was implemented using a different multiplication formula or a combination of multiplication methods. Thus, each partial multiplier has its own gate complexity, structure and circuit. Our experiments shows clearly that the *kP* design with the partial multiplier implemented using the classical multiplication formula only is the most resistant one against the performed DPA attack for both gate technologies.